\documentclass[a4paper,11pt]{article}
\usepackage{pos}
\usepackage{lineno}
\newcommand{\pp}{\ensuremath{\mathrm {p\kern-0.05em p}}}
\newcommand{\PbPb}{\ensuremath{\mbox{Pb--Pb}}}
\newcommand{\GeVc}{\ensuremath{\mathrm{GeV}\kern-0.05em/\kern-0.02em c}}

\newcommand{\pT}{\ensuremath{p_{\mathrm{T}}}}

\newcommand{\kT}{\ensuremath{k_{\mathrm{T}}}}

\newcommand{\pTjet}{\ensuremath{p_{\mathrm{T,\;jet}}}}

\newcommand{\tg}{\ensuremath{\theta_{\mathrm{g}}}}

\newcommand{\zg}{\ensuremath{z_{\mathrm{g}}}}

\newcommand{\ang}{\ensuremath{\lambda_{\alpha}}}
\newcommand{\pTsubjet}{\ensuremath{p_{\mathrm{T}}^{\mathrm{ch\; subjet}}}}
\newcommand{\zr}{\ensuremath{z_{\mathrm{r}}}}

\title{Jet substructure and correlations in hadronic final states from ALICE}
\ShortTitle{Jet substructure and correlations from ALICE}

\author*[a,b]{James Mulligan for the ALICE Collaboration}
\emailAdd{james.mulligan@berkeley.edu}
\affiliation[a]{Nuclear Science Division, Lawrence Berkeley National Laboratory, Berkeley, California 94720, USA}
\affiliation[b]{Physics Department, University of California, Berkeley, CA 94720, USA}

\abstract{Jets of high energy collimated particles provide a rich phenomenology 
to study quantum chromodynamics,
from first-principles tests of perturbative calculations to 
investigations of the emergent properties of the strongly-coupled quark-gluon plasma.
In these proceedings, we highlight several recent jet measurements 
by the ALICE Collaboration, with a focus on jet substructure observables.
}

\FullConference{%
 
 The Ninth Annual Conference on Large Hadron Collider Physics - LHCP2021\\
 7-12 June 2021\\
 Online
}

\begin{document}
\maketitle

\section{Introduction}

Jet observables in both proton-proton and heavy-ion collisions 
can be used to study fundamental aspects of quantum chromodynamics (QCD).
In \pp{} collisions, jet measurements test state-of-the-art perturbative calculations
and explore the transition from the perturbative to the nonperturbative regimes.
In \PbPb{} collisions, jets serve as probes of the quark-gluon plasma (QGP)
to study the physical properties of deconfined QCD matter \cite{ReviewXinNian, ReviewYacine, ReviewMajumder, JETSCAPE:2021ehl}.
In both cases, studying the internal pattern of particles within jets,
known as jet substructure, enables the design of observables to target
specific regions of QCD phase space \cite{Larkoski_2020}.

In these proceedings, we highlight a selection of recent results from the 
ALICE experiment \cite{aliceDetector}, with an emphasis on analytically calculable jet substructure observables. 
All results presented utilize jets reconstructed from charged particles at midrapidity using the
anti-\kT{} algorithm \cite{antikt}, and are corrected for detector effects and (in \PbPb{} collisions) underlying event fluctuations.

\section{Jet measurements in proton--proton collisions}

\begin{figure}[!b]
\centering{}
\includegraphics[scale=0.5]{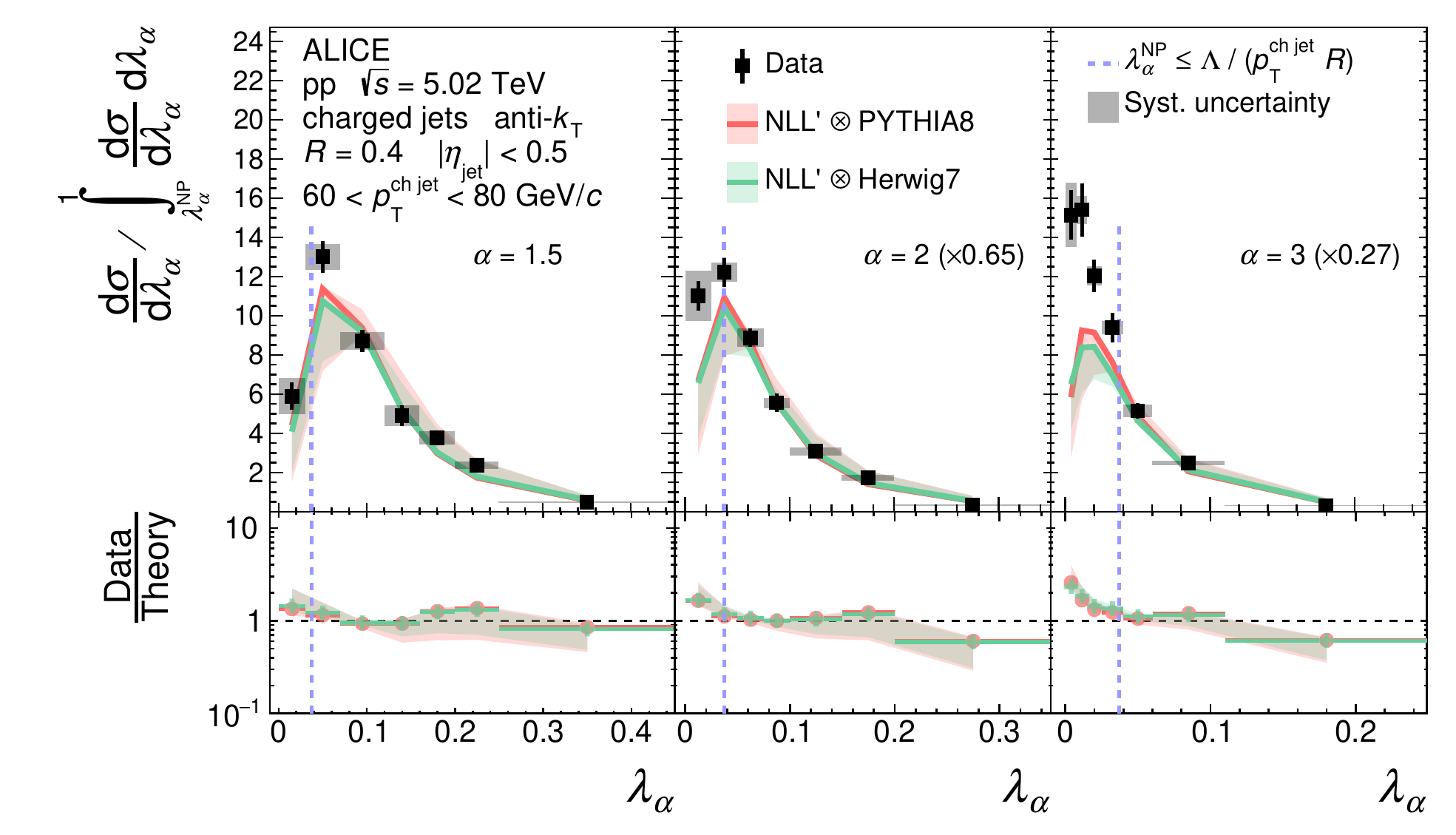}
\caption{Comparison of measured ungroomed jet angularities \ang{} in \pp{}
collisions for $\alpha=1.5,2,3$ to analytical
NLL$^\prime$ predictions \cite{Kang_2018} with MC hadronization and charged particle corrections \cite{ALICE:2021njq}.}
\label{fig:ang}
\end{figure}

\textit{Jet angularities.}
The infrared and collinear safe jet angularities \cite{Almeida:2008yp, Larkoski_2014}
provide a flexible way to study QCD in both \pp{} 
and \PbPb{} \cite{Aad_2012, PhysRevD.98.092014, ang2018} collisions by
systematically varying the weight of collinear radiation via the parameter $\alpha>0$:
\begin{equation} \label{ang_eqn}
\lambda_\alpha =\sum\limits_{i \in \text{jet}}
\bigg( \frac{p_{\text{T},i}}{p_{\text{T, jet}}} \bigg)
\bigg( \frac{\Delta R_i}{R} \bigg)^\alpha.
\end{equation}
ALICE recently measured the ungroomed, and, for the first time, the groomed jet angularities 
in \pp{} collisions for a variety of $\alpha$ \cite{ALICE:2021njq}, 
shown in Fig. \ref{fig:ang}. 
The distributions are compared to SCET calculations 
\cite{Kang_2018, KANG201941} using a Monte Carlo (MC) based folding procedure,
and reveal a transition from generally good agreement in the perturbative regime to deviation in the
nonperturbative regime. 

\begin{figure}[!b]
\centering{}
\includegraphics[scale=0.3]{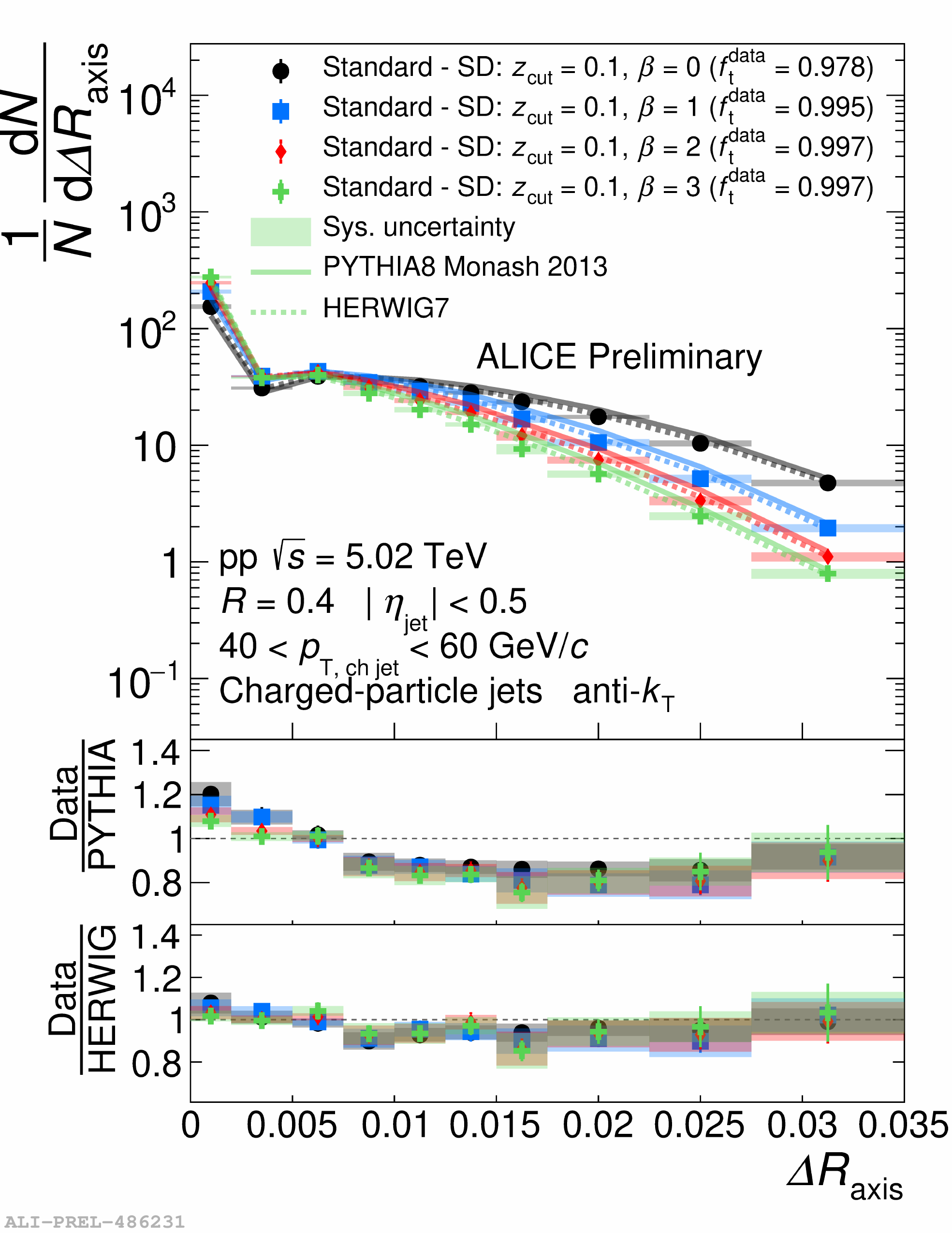}
\qquad
\includegraphics[scale=0.3]{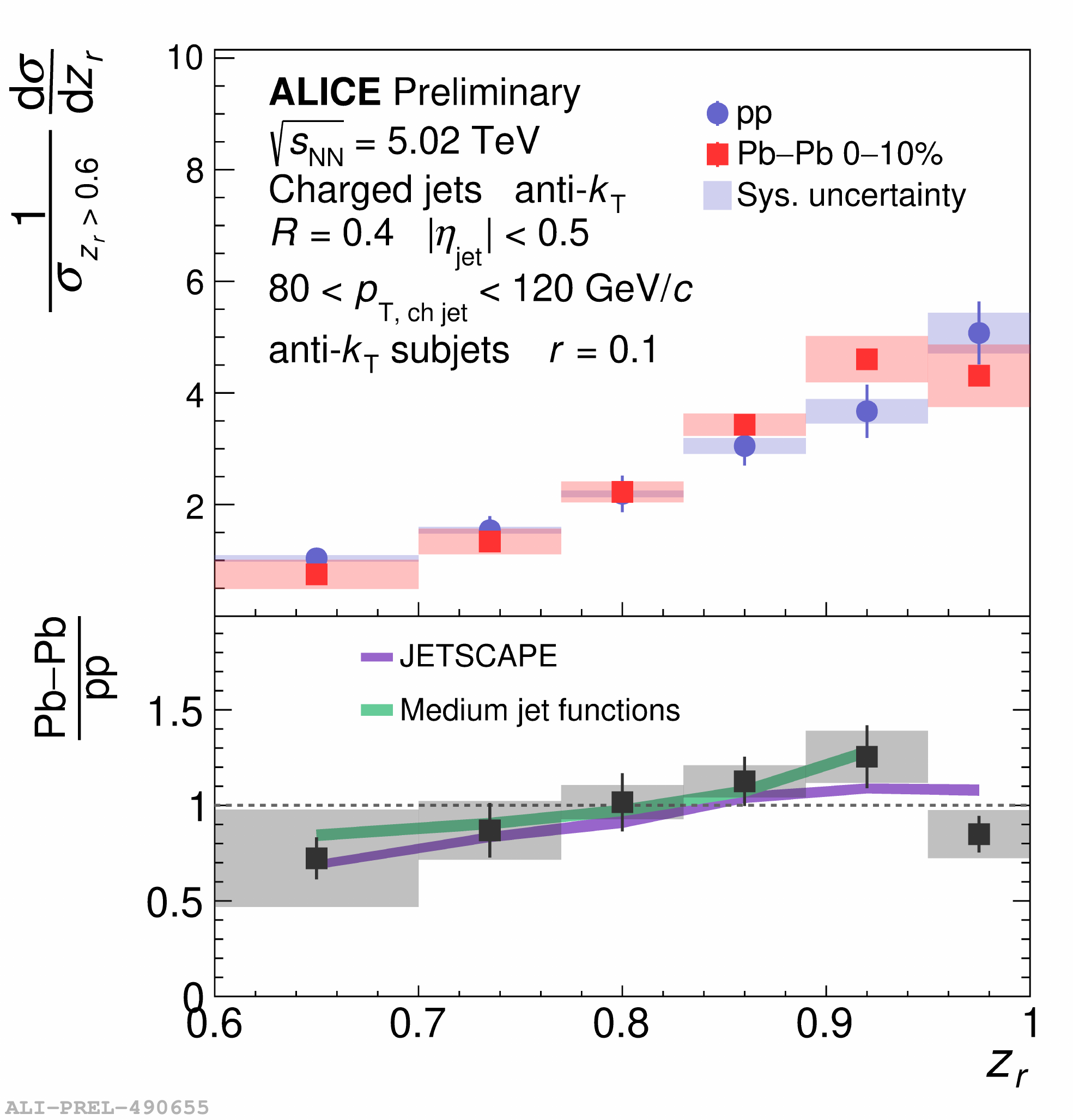}
\caption{
Left: ALICE measurements of pairwise jet axis angular differences for a variety of Soft Drop
grooming conditions, compared to PYTHIA \cite{pythia} and HERWIG \cite{Bellm:2015jjp}.
Right: ALICE measurements of leading subjet fragmentation in \pp{} and \PbPb{} collisions, 
compared to theoretical predictions \cite{Kang:2017mda,Qiu:2019sfj,Putschke:2019yrg, LBT, Majumder_2013}.}
\label{fig:axis}
\end{figure}

\textit{Jet axis differences.}
The soft, wide-angle substructure of jets can be studied by comparing the jet-by-jet rapidity-azimuth 
difference between pairs of jet axes:
\begin{equation} \label{axis_eqn}
\Delta R_{\mathrm{axis}} = \sqrt{\Delta y^2 + \Delta \varphi ^2},
\end{equation}
where the axes can be defined by (i) the standard $E$-scheme recombination axis, 
(ii) the Soft Drop (SD) groomed axis, or the (iii) winner-take-all (WTA) recombination axis \cite{Cal:2019gxa}.
Figure \ref{fig:axis} (left) shows the first measurement of these pairwise axis differences,
where the comparison of the standard and SD axes show small
absolute differences which increase as the grooming condition becomes larger.
These soft-sensitive observables can be used to study a variety of nonperturbative physics \cite{Cal:2019gxa}.

\section{Jet measurements in heavy-ion collisions}

\textit{Subjet fragmentation.}
In heavy-ion collisions, measurements of reclustered subjets have been proposed as
sensitive probes of jet quenching \cite{Kang:2017mda, Neill:2021std, Apolinario:2017qay}.
We consider first inclusively clustering jets with the anti-\kT{} jet algorithm and jet radius $R$,
and then reclustering the jet constituents with the anti-\kT{} jet algorithm and subjet radius $r<R$.
We then consider the fraction of transverse momentum carried by the
subjet compared to the initial jet:
$\zr = \pTsubjet / \pTjet$.
Figure \ref{fig:axis} (right) shows the distribution of leading subjets with $r=0.1,R=0.4$ 
in both \pp{} and \PbPb{} collisions. 
The distributions are compared to theoretical predictions \cite{Kang:2017mda,Qiu:2019sfj,Putschke:2019yrg, LBT, Majumder_2013} which accurately reproduce a mild rising trend of the ratio with \zr{}, which can be attributed to jet collimation, which then falls as $\zr \rightarrow 1$, which may be due to the large
quark/gluon fraction at $\zr \rightarrow 1$.
These measurements offer an opportunity to probe higher $z$ than hadron fragmentation measurements,
and are an important ingredient for future tests of the universality of in-medium jet fragmentation functions.

\begin{figure}[!t]
\centering{}
\includegraphics[scale=0.26]{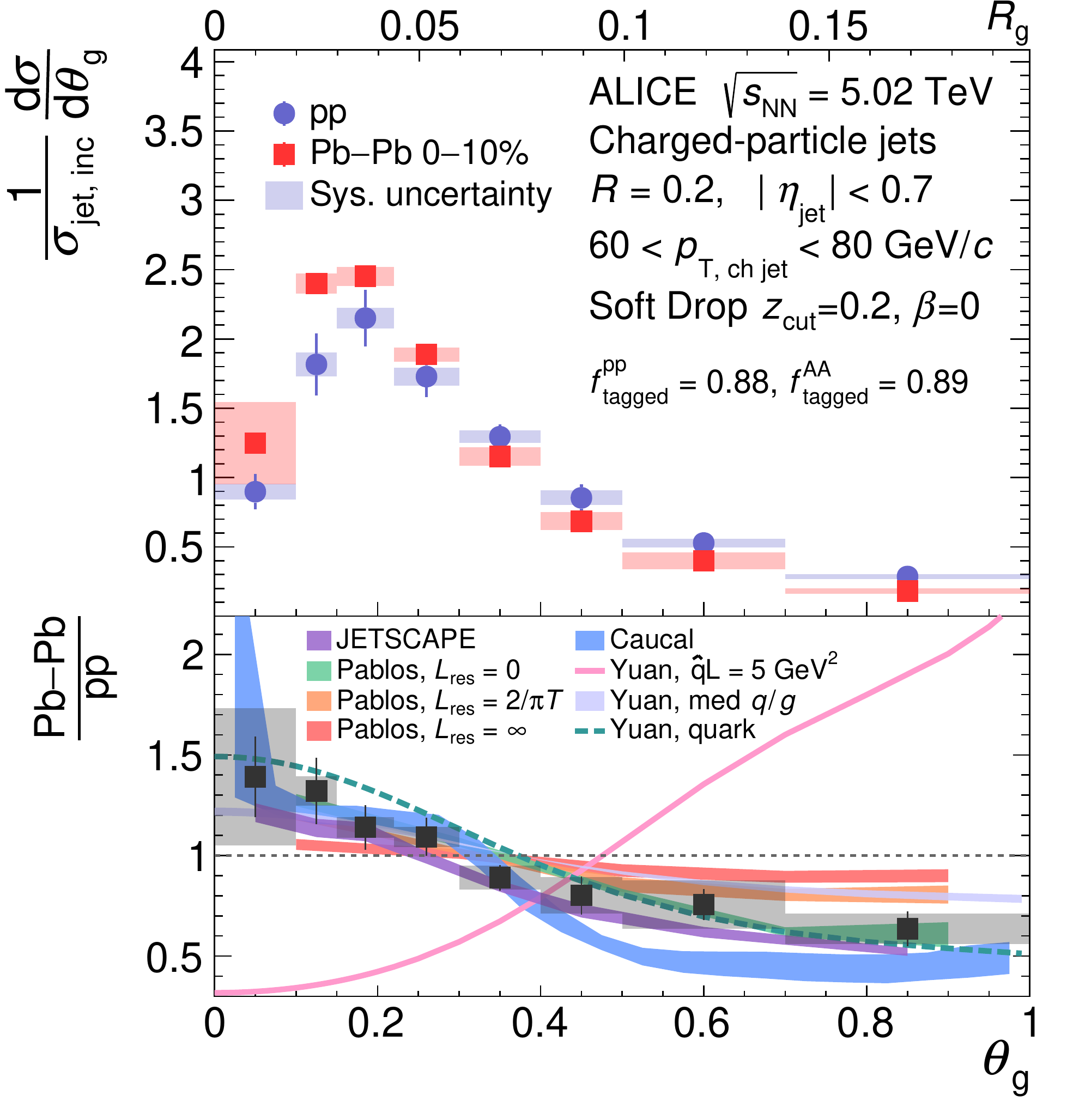}
\qquad
\includegraphics[scale=0.24]{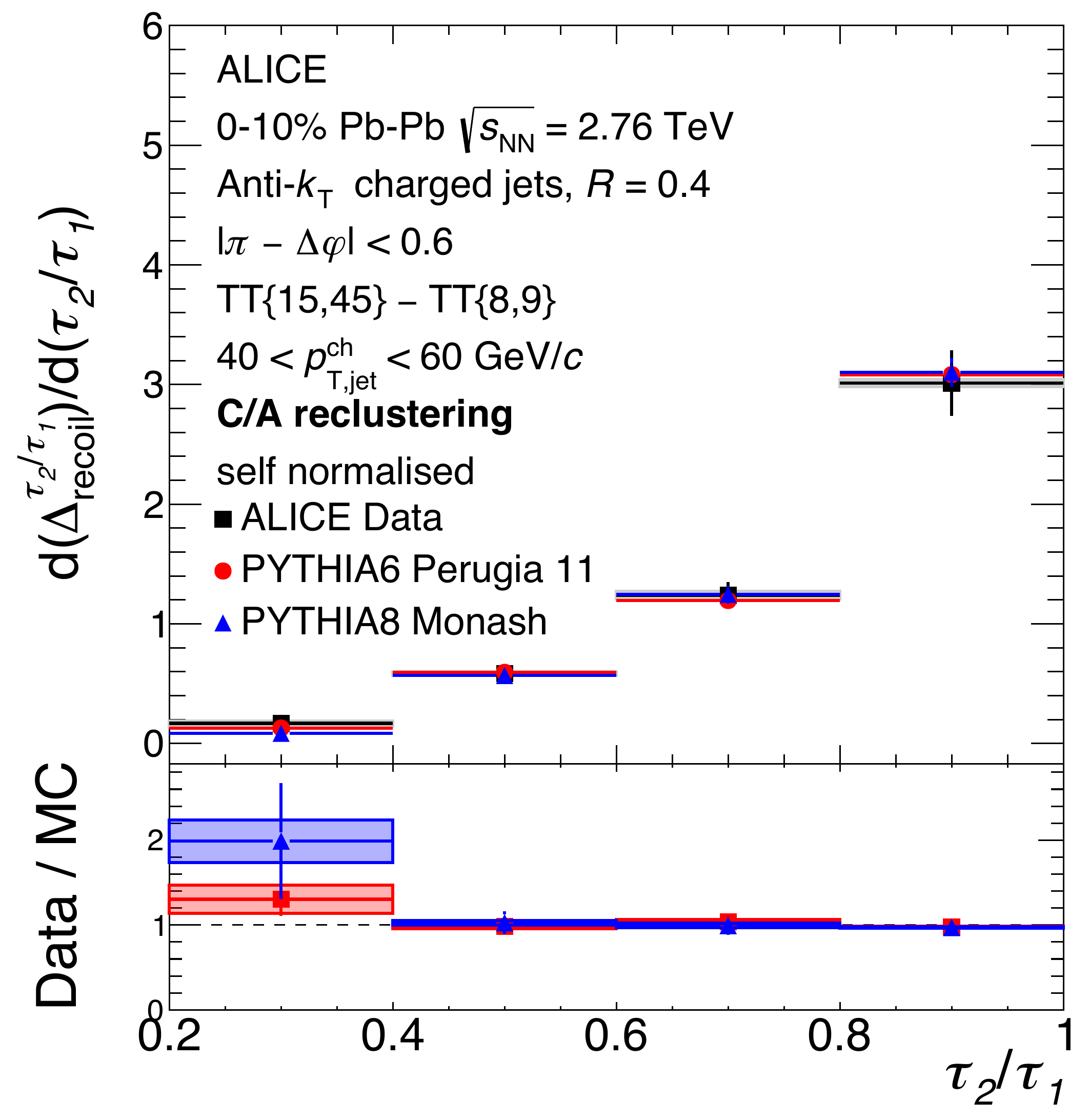}
\caption{Left: Measurements of \tg{} in \PbPb{} compared to \pp{} collisions \cite{ALICE:2021obz}.
Right: Measurements of the $\tau_2/\tau_1$ N-subjettiness distribution in \PbPb{} collisions  \cite{ALICE:2021vrw}
compared to PYTHIA \cite{pythia}.}
\label{fig:sd}
\end{figure}

\textit{Groomed jet radius.}
Jet grooming techniques \cite{Larkoski:2014wba, Dasgupta:2013ihk, Larkoski:2015lea}
have been applied to heavy-ion collisions
to explore whether jet quenching 
modifies the hard substructure of jets
\cite{PhysRevLett.119.112301, Mehtar-Tani2017, Chang:2019nrx, Elayavalli2017, Caucal:2019uvr, Ringer_2020, Casalderrey-Solana:2019ubu, Andrews_2020, PhysRevLett.120.142302, Acharya:2019djg, Sirunyan2018}.
By using strong grooming conditions \cite{Mulligan:2020tim}, ALICE 
measured the groomed momentum fraction, \zg{} \cite{Cal:2021fla},
and the groomed jet radius, \tg{} \cite{Kang:2019prh}, with the Soft Drop
grooming algorithm. 
Figure \ref{fig:sd} (left) shows a narrowing of the \tg{} distributions 
in \PbPb{} collisions relative to  \pp{} collisions \cite{ALICE:2021obz}.
These measurements are compared to a variety of jet quenching models \cite{Putschke:2019yrg, LBT, Majumder_2013, Caucal:2019uvr, Caucal_2018, PhysRevLett.119.112301, Chang:2019nrx, HybridModel, HybridModelResolution, Casalderrey-Solana:2019ubu, Ringer_2020},
most of which capture the qualitative narrowing effect observed. 
This behavior is consistent with models implementing an incoherent interaction of the 
jet shower constituents with the medium, but is also consistent with
medium-modified quark/gluon fractions with fully coherent energy loss --
presenting the opportunity for future measurements to disentangle them definitively.

\textit{N-subjettiness.}
Semi-inclusive hadron-jet correlations are well-suited to 
statistical background subtraction procedures in heavy-ion collisions, which
allows jet measurements to low \pT{} and large $R$ \cite{hjetPbPb, hjetAuAu}.
Recently, ALICE measured the N-subjettiness \cite{Stewart:2010tn,Thaler:2010tr} of jets
recoiling from a high-\pT{} hadron with this method \cite{ALICE:2021vrw}.
Figure \ref{fig:sd} (right) shows the distribution of per-trigger semi-inclusive
yields of the $\tau_2/\tau_1$ ratio in \PbPb{} collisions compared to PYTHIA \cite{pythia},
which show no significant modification in the pronginess of jets in heavy-ion collisions.
This suggests that medium-induced emissions are not sufficiently hard to produce a distinct 
secondary prong, in line with the lack of modification of \zg{} observed \cite{ALICE:2021obz}.
        
\section{Conclusion}

We have presented several new ALICE measurements of jet substructure in \pp{} collisions, 
which provide new tests of our first-principles understanding of QCD by
exploring the transition from perturbative to nonperturbative physics.
In heavy-ion collisions, ALICE measurements are producing an emerging picture of jet quenching phenomenology:
hard splittings are not strongly modified, as evidenced by $\zg, \tau_N$, 
but there is a strong collimation or filtering effect of wide jets, as evidenced by \tg. 
The medium-induced soft splitting responsible for this filtering may be exposed in
regions dominated by quark jets, as suggested by high-\zr{} subjet fragmentation.
Together, these observables offer future prospects to constrain physical properties of the QGP using global analyses.

\clearpage
{
\scriptsize
\bibliographystyle{JHEP}
\bibliography{main.bib}
}

\end{document}